\documentclass[smallextended]{svjour3}       % onecolumn (second format)
\usepackage[T1]{fontenc}
\usepackage{amsmath}
\usepackage{txfonts}
\usepackage{graphicx}

\smartqed  % flush right qed marks, e.g. at end of proof
\newcommand{\ket}[1]{|#1\rangle}
\newcommand{\bra}[1]{\langle #1 |}

\DeclareMathOperator{\Tr}{Tr}

\begin{document}
\hypersetup{pdftitle={Quantum Strongly Secure Ramp Secret Sharing},pdfauthor={Paul Zhang, Ryutaroh Matsumoto},pdfkeywords={quantum secret sharing, non-perfect secret sharing, ramp secret sharing, strong security}}
\journalname{Quantum Information Processing}
\title{Quantum Strongly Secure Ramp Secret Sharing}
\author{Paul Zhang         \and
        Ryutaroh Matsumoto %etc.
}
\institute{Paul Zhang \at
              California Institute of Technology, USA\\
              ORCID: 0000-0003-4136-1315 \\
              \email{pzpzpzp1@gmail.com}           %  \\
%             \emph{Present address:} of F. Author  %  if needed
           \and
           Ryutaroh Matsumoto \at
              Department of Communications and Computer Engineering\\
              Tokyo Institute of Technology, Japan \\
              ORCID: 0000-0002-5085-8879 \\
              \email{ryutaroh@it.ce.titech.ac.jp} 
}

\date{August 8, 2014}
\maketitle
\begin{abstract}
Quantum secret sharing is a scheme for encoding a quantum state (the secret) into multiple shares and distributing them among several participants. If a sufficient number of shares are put together, then the secret can be fully reconstructed. If an insufficient number of shares are put together however, no information about the secret can be revealed. In quantum ramp secret sharing, partial information about the secret is allowed to leak to a set of participants, called an unqualified set, that cannot fully reconstruct the secret. By allowing this, the size of a share can be drastically reduced. This paper introduces a quantum analog of classical strong security in ramp secret sharing schemes. While the ramp secret sharing scheme still leaks partial information about the secret to unqualified sets of participants, the strong security condition ensures that qudits with critical information can no longer be leaked. 
\end{abstract}
 
\section{Introduction}
\label{sec1}
Secret sharing (SS) \cite{shamir79} is a cryptographic scheme to
encode a secret to multiple shares being distributed to
participants, so that only \emph{qualified sets} of participants
can reconstruct the original secret from their shares.
Traditionally both secret and shares were classical information
(bits). Several authors \cite{cleve99,gottesman00,smith00}
extended the traditional SS to a quantum one
so that a quantum secret is encoded to quantum shares.

SS can be classified into two categories.
One is perfect SS and the other is non-perfect or ramp SS
\cite{ogata93}, \cite[Chapter 13]{stinson06}.
In perfect SS,
we require every unqualified set of participants
to have zero information of the secret,
while in non-perfect SS we do not require such a property.
A major disadvantage of perfect SS is that
the size of each share must be larger than or equal to
that of secret, in both classical case \cite{capocelli93}
and quantum case \cite{cleve99,gottesman00,smith00}.
By tolerating partial information leakage to
unqualified sets, the size of shares can be much smaller
than that of secret. Such an SS is called a ramp SS or
a non-perfect SS
\cite{blakley85,iwamoto06,ogata93,yamamoto86}. The quantum ramp SS was
proposed by Ogawa et al.\ \cite{ogawa05}.
In both Ogawa et al.'s scheme and our proposal,
the size of shares is $L$ times smaller than that of the secret,
where $L$ is the number of qudits in a secret.
This paper focuses only on the quantum ramp SS.

For a general ramp SS, an unqualified set of participants
is allowed to have partial information of the secret. This
can be undesirable in some cases.
For example,
consider a classical secret representing $``username:password$''.
When an unqualified set
has $8$-symbol partial information of the secret,
it could know all of $``password$'', which is critical confidential information.
The existence of such a case
is demonstrated in an explicit example by
Iwamoto and Yamamoto \cite{iwamoto06}.
The first purpose of this paper is to demonstrate
a similar danger in the quantum ramp SS \cite{ogawa05}
by providing
an explicit example.

In order to prevent such cases,
Yamamoto \cite{iwamoto06,yamamoto86} defined the strong
security of classical ramp SS as follows:
A $(k$, $L$, $n)$ strongly secure classical ramp SS
distributes $L$ symbols of a classical secret $\vec{s}$
to $n$ symbols such that any $k$ or more participants
can determine $\vec{s}$ while any $i$ symbols of $\vec{s}$ is
kept completely secret to any $(k-i)$ or less participants, where each participant has one symbol.
Thus a strongly secure ramp SS excludes the danger explained
in the last paragraph.

The second purpose of this paper
is to adapt the strong security criterion in the classical
case to the quantum case.
We define 
a quantum
$(k$, $L$, $n)$ 
strongly secure ramp SS
as a quantum ramp SS
that
distributes $L$ qudits of a quantum secret $\sigma$
to $n$ qudits $\rho$ such that
any $k$ or more participants
can determine $\sigma$ while any $i$ qudits of $\sigma$ is
kept completely secret from any $k-i$
or less participants, where each participant has one qudit.

The third purpose of this paper is to
provide an explicit construction of quantum ramp SS
realizing the strong security with the same efficiency as
the conventional quantum ramp SS in \cite{ogawa05}.
The difference of constructions between \cite{ogawa05} and ours is as
follows:
While Ogawa et al.\ \cite{ogawa05}
encode a secret to coefficients of a polynomial,
we encode it to the function values of a polynomial
as done in \cite{mceliece81,nishiara09} for the construction of
classical strongly secure SS.

We stress that this paper studies ramp (non-perfect) SS
while \cite{cleve99,gottesman00,smith00} studied
perfect SS, and that none of the results in this paper
are contained in \cite{cleve99,gottesman00,smith00}.

This paper is organized as follows:
Section \ref{secLSS} demonstrates how Ogawa et al.'s encoding fails to satisfy
an intuitive quantum version of Yamamoto's classical strong security.
Section \ref{sec2} formalizes the strong security criterion
for quantum ramp SS.
Section \ref{sec31} proposes the encoding of secrets.
Section \ref{sec3} proposes the decoding of secrets.
Section \ref{sec4} proves the strong security of
the proposed scheme.
Section \ref{sec6} gives concluding remarks.

\section{Motivation for Quantum Strong Security}
\label{secLSS}

We will show here how Ogawa et al.'s secret sharing scheme fails to satisfy the strong security condition.
Let $\mathcal{G}_i$ and $\mathcal{H}_j$ be $q$-dimensional complex linear spaces.
We refer to quantum systems represented by $\mathcal{G}_i$ and $\mathcal{H}_j$
as \emph{qudits}.
We assume $q$ to be a prime power, and denote
by $\mathbf{F}_q$ the finite field
with $q$ elements.
We also assume that orthonormal bases of $\mathcal{G}_i$
and $\mathcal{H}_j$ are indexed by $\mathbf{F}_q$
as $\{\ket{s}\}_{s \in \mathbf{F}_q}$.

In Ogawa et al.'s $(k$, $L$, $n)$ ramp secret sharing scheme, the encoding of a quantum secret is defined by unique public values $(x_1, \ldots, x_n) \in \mathbf{F}_q^n$ and the transformation of an $L$-qudit secret $\ket{s_1, \ldots, s_L}$, to 
\begin{align*}
\frac{1}{\sqrt{q^{k-L}}}\sum_{\vec{c}\in D(s^L)}
\ket{p_{\vec{c}}(x_1),\dots,p_{\vec{c}}(x_n)},
\label{kLn-const-4}
\end{align*}
where
$
D(s^L)=\left\{(c_1,\ldots,c_k)\in \mathbf{F}^k_q \mid \forall_{i \in \left\{ 1,\ldots,L\right\}} \; c_i=s_i\right\},
$ and 
$ p_{\vec{c}}(x)=c_1+c_2x+\ldots+c_{k-1}x^{k-1}
$
 the polynomial whose coeffcients are specified by $\vec{c}=(c_1, \ldots, c_k)$. We now provide a specific example in which an unqualified set is able to reconstruct one qudit of the secret.

\begin{example}
\label{example1}
We consider this scheme where
the number of qudits in the secret $L=2$, the minimal number of participants needed to decode the secret $k=3$, the number of shares $n=4$, and the size of the field $q=7$.
Public values $\vec{x}=(2,3,1,6) \in \mathbf{F}_7^4$.
The quantum secret has $q^L = 49$ dimensions.
Its orthonormal basis is the set
$\{ \ket{s_1 s_2} \mid s_1, s_2 \in \mathbf{F}_7\}.
$

We choose a particular basis state $\ket{s_1 s_2}$ to be the secret and
consider the shares encoded from $\ket{s_1 s_2}$.
The set $D(s_1, s_2)$ consists of coefficients $\vec{c}_i=(s_1, s_2, r_i)$ 
where 
$i\in\left\{1,\ldots,7\right\}$ and $r_i=i-1 \in \mathbf{F}_7$.
These coefficients specify polynomials to be in the form: 
\[
p(x)_i = s_1 + s_2 x + r_i x^2.
\]
The four shares, corresponding to $\vec{c}_i=(s_1, s_2, r_i)$, are therefore given by
\begin{eqnarray*}
p_{\vec{c}_i}(2) &=& s_1 + 2s_2 + 4r_i,\\
p_{\vec{c}_i}(3) &=& s_1 + 3s_2 + 2r_i,\\
p_{\vec{c}_i}(1) &=& s_1 + s_2 + r_i,\\
p_{\vec{c}_i}(6) &=& s_1 + 6s_2 + r_i.
\end{eqnarray*}
The encoded state of four shares is 
\begin{align*}
\frac{1}{\sqrt{7}} (
\ket{p_{\vec{c}_1}(2),p_{\vec{c}_1}(3),p_{\vec{c}_1}(1),p_{\vec{c}_1}(6)}&+
\ket{p_{\vec{c}_2}(2),p_{\vec{c}_2}(3),p_{\vec{c}_2}(1),p_{\vec{c}_2}(6)}+\\
\ket{p_{\vec{c}_3}(2),p_{\vec{c}_3}(3),p_{\vec{c}_3}(1),p_{\vec{c}_3}(6)}&+
\ket{p_{\vec{c}_4}(2),p_{\vec{c}_4}(3),p_{\vec{c}_4}(1),p_{\vec{c}_4}(6)}+\\
\ket{p_{\vec{c}_5}(2),p_{\vec{c}_5}(3),p_{\vec{c}_5}(1),p_{\vec{c}_5}(6)}&+
\ket{p_{\vec{c}_6}(2),p_{\vec{c}_6}(3),p_{\vec{c}_6}(1),p_{\vec{c}_6}(6)}+\\
\ket{p_{\vec{c}_7}(2),p_{\vec{c}_7}(3),p_{\vec{c}_7}(1),p_{\vec{c}_7}(6)}& 
).
\end{align*}
Since $1^2 = (-1)^2 = 6^2 = 1 \in \mathbf{F}_7$,
the coefficients of $r_i$ in
the 3rd and the 4th shares are both $1$,
which provides us with the classical equation
\begin{equation}
s_2 = 4 p(1)_i - 4 p(6)_i. \label{eq1}
\end{equation}
This shows that the value $s_2$ can be retrieved using information from the 3rd and 4th participants. In the quantum setting, the 3rd  and the 4th participants can collectively
apply a unitary matrix $\mathcal{U}$ based on \eqref{eq1}
to produce $\ket{s_2}$ (entangled with the rest of the shares). 

A full rank matrix that performs our desired classical transformation is
\begin{align}
\nonumber
M=
\left(
\begin{array}{cc}
4&4 \\
-4&4
\end{array}
\right)
=
\left(
\begin{array}{cc}
4&4 \\
3&4
\end{array}
\right).
\end{align}
This classical transformation performs a change of basis indices for our quantum state. The unitary quantum transformation $\mathcal{U}$, corresponding to $M$,
sends $\ket{p_{\vec{c}_i}(1), p_{\vec{c}_i}(6)}$ of the $3rd$ and $4th$ participants to 
\[\ket{(p_{\vec{c}_i}(1), p_{\vec{c}_i}(6))\cdot M}=\ket{s_2, s_1+r_i}. \]
$\mathcal{U}$ is unitary because it just permutes basis vectors.

After applying such a transformation on the 3rd and 4th shares, a measurement in the
basis $\{\ket{0}$, \ldots, $\ket{6}\}$ can read off the index $s_2$.
If the $s_2$-component of the quantum secret happens to be
classical information, then it can be completely read off
by the 3rd and the 4th participants. Therefore two shares (an unqualified set) are capable of obtaining partial information of the secret. The strong security condition exists to make sure that no qudits of critical information can be revealed to an unqualified set of participants. Therefore, Ogawa et al.'s encoding is not a strongly secure ramp secret sharing scheme.

Also observe that the 3rd and the 4th participants
can figure out $s_2$ even when the first qudit $\ket{s_1}\bra{s_1}$
is the fully mixed state $I_{7 \times 7} / 7$ as in Definition \ref{def1}.
\end{example}

\section{Definition of the Strong Security}
\label{sec2}
To formally define quantum strong security, we use the same $\mathcal{G}_i$ and $\mathcal{H}_j$ as earlier.
\begin{definition} 
\label{def1}
For integers $0 < L < k < n$,
we define a quantum $(k,L,n)$-threshold strongly secure
ramp secret sharing scheme as
a completely positive trace-preserving map \cite[Chapter 8]{chuangnielsen}
$W$ of $\sigma$
on the state space $\mathcal{S}(\bigotimes_{i=1}^L \mathcal{G}_i)$
into $\mathcal{S}(\bigotimes_{j=1}^n \mathcal{H}_j)$
with the following conditions.
\begin{enumerate}
\item
\label{l1} $\sigma$ can be reconstructed from
any $k$ or more qudits of $\mathcal{H}_j$.
\item
\label{l2} Let $\mathcal{I} \subseteq \{1$, \ldots, $L\}$,
$\overline{\mathcal{I}} = \{1$, \ldots, $L\} \setminus \mathcal{I}$,
$\sigma_1 \in \mathcal{S}(\bigotimes_{i \in \mathcal{I}}\mathcal{G}_i)$,
$\rho_{\mathrm{mix}, \overline{\mathcal{I}}}$ 
the fully mixed state in $\mathcal{S}(\bigotimes_{i \in \overline{\mathcal{I}}}\mathcal{G}_i)$,
$\mathcal{J} \subset \{1$, \ldots, $n\}$, and
$\overline{\mathcal{J}} = \{1$, \ldots, $n\} \setminus \mathcal{J}$.
If $|\mathcal{J}| \leq k-|I|$,  then
$\mathrm{Tr}_{\bigotimes_{j \in \overline{\mathcal{J}}}\mathcal{H}_j}W(\sigma_1 \otimes
\rho_{\mathrm{mix}, \overline{\mathcal{I}}})$ is independent of $\sigma_1$.
In other words, no quantum information of $\sigma_1$ 
is leaked to shares whose indices belong to $\mathcal{J}$.
\end{enumerate}
\end{definition}

Observe that $\sigma$ represents a quantum secret and
$n$ qudits in $\mathcal{S}(\bigotimes_{j=1}^n \mathcal{H}_j)$ represent $n$ shares
distributed to $n$ participants.
Note, Condition \ref{l2} with $I=\{1$, \ldots, $L\}$
is equivalent to the conventional security definition
\cite[Definition 1]{ogawa05}.
In Condition \ref{l2},
we assume that the unqualified set of participants
represented by $\mathcal{J}$ has no prior knowledge on the
quantum state on $\overline{\mathcal{I}}$.
This lack of prior knowledge is expressed as
the fully mixed state $\rho_{\mathrm{mix}, \overline{\mathcal{I}}}$
in $\mathcal{S}(\bigotimes_{i \in \overline{\mathcal{I}}}\mathcal{G}_i)$.

When the quantum secret is an output of a nearly optimal
quantum data compression
\cite[Section 12.2.2]{chuangnielsen}, then
the output is close to the fully mixed state,
otherwise it allows further compression.
The fully mixed state in 
Condition \ref{l2} is also justified in such a case.

\section{Encoding Secrets}
\label{sec31}
We will propose an explicit construction satisfying the
conditions of the last section. Our proposal is a quantum
version of classical strongly secure secret sharing \cite{mceliece81,nishiara09}.
As in \cite{ogawa05} we set $n=2k - L$ indicating a pure state quantum secret sharing scheme.

Let $\mathbf{F}_q$ be as it was previously.
Let $x_1$, \ldots, $x_L$, $y_1$, \ldots, $y_n$
denote publicly known pairwise distinct
elements in $\mathbf{F}_q$.
Let $D_k$ be equivalent to $\mathbf{F}_q^k$ with an interpretation as the set of coefficients of univariate polynomials over $\mathbf{F}_q$
with degree less than $k$.
For $\vec{u} = (u_1$, \ldots, $u_\ell) \in \mathbf{F}_q^\ell$
with pairwise distinct $u_1$, \ldots, $u_\ell$ and
$\vec{c} \in D_k$, define
$P_{\vec{c}}(\vec{u}) = (p_{\vec{c}}(u_1)$, \ldots, $p_{\vec{c}}(u_\ell))
\in \mathbf{F}_q^\ell$ 
to be the evaluation of a polynomial specified by $\vec{c}$, at points $u_1, \ldots, u_\ell$. 

Ogawa et al.\ \cite[Lemma 3]{ogawa05} proved the following lemma:
\begin{lemma}
\label{lem1}

Given a vector $\vec{A}\in\mathbf{F}_q^m$ whose elements are pairwise unique, 
the map $\vec{c} \in D_k \mapsto P_{\vec{c}}(\vec{A})
\in \mathbf{F}_q^m$ is injective if $m \ge k$, and it is bijective if $m=k.$
\end{lemma}

Denote $(x_1$, \ldots, $x_L)$ by $\vec{x}$ and
$(y_1$, \ldots, $y_n)$ by $\vec{y}$,
and let $D_k(\vec{s}) = \{ \vec{c} \in D_k \mid
P_{\vec{c}}(\vec{x}) = \vec{s}\}$ for $\vec{s} \in \mathbf{F}_q^L$.
By Lemma \ref{lem1} we see $|D_k(\vec{s})| = q^{k-L}$.
Consider the encoding map $W$ sending
$\ket{s_1, s_2, \ldots, s_L} \in \bigotimes_{i=1}^L \mathcal{G}_i$
to 

\begin{align}
\frac{1}{\sqrt{q^{k-L}}}
\sum_{\vec{c} \in D_k(s_1, \ldots, s_L)}
\ket{P_{\vec{c}}(\vec{y})}\in \bigotimes_{j=1}^n \mathcal{H}_j,
\label{encoding}
\end{align}
where $s_i \in \mathbf{F}_q$.
This map embeds an orthonormal basis of 
$\bigotimes_{i=1}^L \mathcal{G}_i$ into that of $\bigotimes_{j=1}^n \mathcal{H}_j$.
So this map can be uniquely extended to a complex linear isometric
embedding from $\bigotimes_{i=1}^L \mathcal{G}_i$ into $\bigotimes_{j=1}^n \mathcal{H}_j$.

Observe that the sizes of a secret and shares are the same as
those of the conventional quantum ramp secret sharing
scheme by Ogawa et al.\ \cite{ogawa05},
which means the coding rate of our proposal is
also optimal in the sense of \cite[Corollary 2]{ogawa05}.

\begin{example}
\label{sec32}
As a concrete case, we use the following $(k=3,L=2,n=4)$ quantum ramp secret sharing scheme to encode a secret. Let $\ket{\vec{s}}=\ket{1,5}$ be the secret that we wish to encode. We set the publicly known values $\vec{x}$ and $\vec{y}$ to be $\vec{x}=(1,3)$, and $\vec{y}=(6,2,4,5)$. The set $D_3(1,5)$ can now be calculated explicitly as
\begin{align}
D_3(1,5) =
\left\{
\substack{(6,2,0),(2,5,1),(5,1,2),\\
(1,4,3),(4,0,4),(0,3,5),(3,6,6)}
\right\}.
\label{Dk15}
\end{align}
One can verify that each element $\vec{c}$ of $D_3(1,5)$ has the property that $s_i=c_1+c_2x_i+c_3x_i^2$, as well as the fact that no other polynomials have the same property. 

The encoded secret is the superposition of states indexed by evaluations of polynomials specified by $\vec{c}$, on $\vec{y}.$ We show explicitly that the state corresponding to $\vec{c}=(2,5,1)$, is $\ket{5,2,3,3}$ by the following calculation.
\begin{align}
\nonumber
2+5y_1+y_1^2&=2+5*6+6^2=2-5+1&=5 \\\nonumber
2+5y_2+y_2^2&=2+5*2+2^2=16&=2 \\\nonumber
2+5y_3+y_3^2&=2+5*4+4^2=2+20+16=38&=3 \\\nonumber
2+5y_4+y_4^2&=2+5*5+5^2=2+25+25&=3
\end{align}
One can verify that the states, corresponding to all $\vec{c}\in D_k(1,5)$, in the same order as presented in \eqref{Dk15}, are 
\begin{align}
\nonumber
&\ket{4,3,0,2}, 
\ket{5,2,3,3}, 
\ket{6,1,6,4}, \\\nonumber
&\ket{0,0,2,5}, 
\ket{1,6,5,6}, 
\ket{2,5,1,0}, 
\ket{3,4,4,1}.
\end{align}
The final encoded state is therefore given by
\begin{align}
\nonumber
\ket{\psi_{ex}}=\frac{1}{\sqrt{7}} (
&\ket{4,3,0,2}+ 
\ket{5,2,3,3}+ 
\ket{6,1,6,4}+ \\
\label{example-encode}
&\ket{0,0,2,5}+ 
\ket{1,6,5,6}+ 
\ket{2,5,1,0}+ 
\ket{3,4,4,1}).
\end{align}
Note that if we looked at the subsystem on the last qudit, we have only a fully mixed state. In general, if we looked at any $n-k$ qudit subsystem of the encoded state, we would only find a fully mixed state. We will prove this important fact as \eqref{edecode2} in Section \ref{sec3}.
\end{example}

\section{Decoding Secrets}
\label{sec3}
In this section we will show that by performing the unitary transformations
in \cite[Section VII]{ogawa05} with suitable modifications,
$k$ or more participants can decode the quantum secret $\ket{\vec{s}=(s_1,\ldots,s_L)}$.
As \cite[Section VII]{ogawa05},
we assume that the number of participants is exactly $k$.

Let the notation $A_B$, where $A\in\mathbf{F}_q^m$ is a vector and $B \subseteq \{1, \ldots, m\}$ is an ordered ascending set, be $(a_{b_1}, a_{b_2}, \ldots, a_{b_{|B|}})$ the vector $A$ indexed by $B$. 
Let $\mathcal{J} \subset \{1$, \ldots, $n\}$, where $|\mathcal{J}|=k$, be the set indexing the shares available to the $k$ participants, and
$\overline{\mathcal{J}} = \{1$, \ldots, $n\} \setminus \mathcal{J}$. Now we introduce slightly modified notation from \cite{ogawa05}: 
\[
M_d^c(A_B) =
\left( \begin{array}{ccc}
a_{b_1}^c & \dots & a_{b_{|B|}}^c \\
a_{b_1}^{c+1} & \dots & a_{b_{|B|}}^{c+1} \\
\vdots & & \vdots \\
a_{b_1}^d & \dots & a_{b_{|B|}}^d
\end{array}\right)
\quad{\left(c<d\right)}.
\]
This is generally a submatrix of a Vandermonde matrix. 
Note that $M_{k-1}^0(A_B)$ acts as the linear transformation between the coefficients $\vec{c}$ of a polynomial of degree $k-1$ to its evaluations on the set $A_B$, that is,
\begin{align}
\nonumber
\vec{c} \cdot M_{k-1}^0(A_B) &= \vec{c} 
\left( \begin{array}{ccc}
a_{b_1}^0 & \dots & a_{b_{|B|}}^0 \\
a_{b_1}^1 & \dots & a_{b_{|B|}}^1 \\
\vdots & & \vdots \\
a_{b_1}^{k-1} & \dots & a_{b_{|B|}}^{k-1}
\end{array}\right)
\\\nonumber
&=
\left( \begin{array}{ccc}
c_0a_{b_1}^0 & \dots & c_0a_{b_{|B|}}^0 \\
+&&+\\
c_1a_{b_1}^1 & \dots & c_1a_{b_{|B|}}^1 \\
+&&+\\
\vdots & & \vdots \\
+&&+\\
c_{k-1}a_{b_1}^{k-1} & \dots & c_{k-1}a_{b_{|B|}}^{k-1}
\end{array}\right)
\\\nonumber
&=
\left( p_{\vec{c}}(a_{b_1}), \ldots, p_{\vec{c}}(a_{b_{|B|}}) \right)
\\\nonumber
&=
P_{\vec{c}}(A_B).
\end{align} 
This also means, by Lemma \ref{lem1}, that, if $|A_B| = k$, the matrix $M_{k-1}^0(A_B)$ can be inverted to retrieve coefficients $\vec{c}$ from $P_{\vec{c}}(A_B)$.

As \cite{ogawa05}, without loss of generality, we let $\mathcal{J}=\{1,\ldots,k\}$. Starting with the encoded state \eqref{encoding}, we apply the unitary corresponding (review Example \ref{example1} for the correspondence) to the first decoding matrix
\begin{align}
\label{first-decoding-mat}
M_{k-1}^0(\vec{y}_{\mathcal{J}})^{-1}
\end{align}
on shares indexed by $\mathcal{J}$. As explained above, this transformation will retrieve the polynomial coefficients $\vec{c}$ from $P_{\vec{c}}(\vec{y}_\mathcal{J})$.
The resulting state is
\begin{align}
\frac{1}{\sqrt{q^{k-L}}}
\sum_{\vec{c} \in D_k(\vec{s})}
\ket{\vec{c},P_{\vec{c}}(\vec{y}_{\overline{\mathcal{J}}})}
\label{partdecode}.
\end{align}
Next apply the unitary transformation corresponding to the second decoding matrix:
\begin{align}
\left( 
M_{k-1}^0(\vec{x}) \quad  M_{k-1}^0(\vec{y_{\overline{\mathcal{J}}}}) 
\right)
\label{MMdecode}.
\end{align}
Note that since $n-k=k-L$, the width of \eqref{MMdecode} is $|\vec{x}|+|\vec{y_{\overline{\mathcal{J}}}}|=(L)+(n-k)=L+k-L=k$. This matrix has the effect of taking coefficients $\vec{c}$ and evaluating them at $\vec{x}$ and $\vec{y_{\overline{\mathcal{J}}}}$. By our definition of $D_k(\vec{s})$, the evaluation of the coefficients at $\vec{x}$, is equivalent to the secret. Therefore, the transformation \eqref{MMdecode} takes the state in \eqref{partdecode} to the final state:
\begin{align}
\frac{1}{\sqrt{q^{k-L}}}\ket{\vec{s}} \otimes
\sum_{\vec{c} \in D_k(\vec{s})}
\ket{P_{\vec{c}}(\vec{y}_{\overline{\mathcal{J}}}),P_{\vec{c}}(\vec{y}_{\overline{\mathcal{J}}})}.
\label{edecode}
\end{align}

It remains to verify that the $n-L$ shares left over do not in fact rely on the decoded secret at all. This step is necessary because, if there is any dependency between the decoded secret $\ket{\vec{s}}$ and the remaining $n-L$ shares, the decoded secret will no longer be unentangled from the remaining shares when the secret is a superposition of two different pure states $\ket{\vec{s}_1}$ and $\ket{\vec{s}_2}$. For example,
if the decoding on the encoded pure state secret $\ket{\vec{s}_1}$ produced a dependence by the shares, on the secret, $\ket{\vec{s}_1, f(\vec{s}_1)}$, 
then the decoding of shares encoded from $\frac{1}{\sqrt{2}}(\ket{\vec{s}_1}+\ket{\vec{s}_2})$ will be 
\begin{align}
\frac{1}{\sqrt{2}}
(\ket{\vec{s}_1, f(\vec{s}_1)}+\ket{\vec{s}_2, f(\vec{s}_2)}),
\end{align}
which has a different density operator from the one we desire. We need the decoded secret to be unentangled with the remaining shares, resulting in a state
\begin{align}
\label{separated}
\frac{1}{2}(\ket{\vec{s}_1}+\ket{\vec{s}_2})
(\bra{\vec{s}_1}+\bra{\vec{s}_2})
\otimes \rho,
\end{align}
where $\rho$ is the density operator of the remaining $n-L$ shares. 
Thus to show our decoder successfully decodes the secret, we will prove that the quantum state of the remaining shares have no dependence on the secret.

First we prove that for a set $X\subseteq \mathbf{F}_q
\setminus \{x_1$, \ldots, $x_L\}$ the map $\vec{c} \in D_k(\vec{s}) \rightarrow P_{\vec{c}}(X)$, where $|X|=k-L$, is bijective. This can be seen in the forward direction by the fact that $\vec{c}$ uniquely determines the polynomial's evaluations. In the reverse direction, $\vec{s}$ provides us with $L$ evaluations of the polynomial that, when coupled with $|X|$ more evaluations at distinct points, provides us with $k$ evaluations of the polynomial. By applying Lemma \ref{lem1} the $k$ evaluations uniquely determine the coefficients. Therefore the map between $\vec{c} \in D_k(\vec{s})$ and $P_{\vec{c}}(X)$ for a fixed $\vec{s}\in \mathbf{F}_q^L$ is bijective. 

We now replace the set $X$ with $\vec{y}_{\overline{\mathcal{J}}}$. Since $|D_k(\vec{s})|=q^{k-L}$, \eqref{edecode} can now be re-written as
\begin{align}
\label{edecode2}
\frac{1}{\sqrt{q^{k-L}}}\ket{\vec{s}} \otimes
\sum_{\vec{v} \in \mathbf{F}_q^{k-L}}
\ket{\vec{v}}\ket{\vec{v}}.
\end{align}
%NOTE THE N-L BELOW IS CORRECT. NOT N-K.
The remaning $n-L$ shares are therefore a purification of the fully mixed state on the $\overline{\mathcal{J}}$ subsystem. Since it is unentangled with the first $L$ shares, we have reconstructed the secret.

\begin{example}
\label{example-decode}
We will now provide a concrete example of the decoding scheme acting on our previously calculated encoded state $\ket{\psi_{ex}}$ in \eqref{example-encode}. Recall that the secret we encoded was $\ket{1,5}.$
First we apply the unitary corresponding to $M_{k-1}^0(Y_\mathcal{J})^{-1}$. The matrix $M_{k-1}^0(Y_\mathcal{J})$ and its inverse are shown below.
\begin{align}
\label{M1f}
M_2^0(6,2,4) = \left(
\begin{array}{ccc}
1&1&1\\
6&2&4\\
6^2&2^2&4^2
\end{array}\right)
=\left(
\begin{array}{ccc}
1&1&1\\
6&2&4\\
1&4&2
\end{array}\right)
\end{align}

\begin{align}
\label{M1b}
M_2^0(6,2,4)^{-1} = \left(
\begin{array}{ccc}
1&1&1\\
3&4&1\\
4&2&5
\end{array}\right)
\end{align}
One can verify that the product of (\ref{M1f}) and (\ref{M1b}) is the identity. 

We explicitly show that the result of (\ref{M1b}) acting on state $\ket{4,3,0,2}$ results in $\ket{6,2,0,2}$.
\begin{align}
\nonumber
(4,3,0)
\left(\begin{array}{ccc}
1&1&1\\
3&4&1\\
4&2&5
\end{array}\right)
=
(4+9+0,4+12+0,4+3)=(6,2,0)
\end{align}
From Example \ref{sec32}, one can see that ($\vec{c}, P_{\vec{c}}(\vec{y}_{\overline{\mathcal{J}}})) = (6,2,0,2)$. 
One can verify (\ref{M1b}) acts similarly for the rest of the terms in \eqref{example-encode}, resulting in the following partially decoded state:
\begin{align}
\nonumber
\ket{\psi_{pd}}=\frac{1}{\sqrt{7}} (
&\ket{6,2,0,2}+ 
\ket{2,5,1,3}+ 
\ket{5,1,2,4}+ \\
\label{example-encode-pd}
&\ket{1,4,3,5}+ 
\ket{4,0,4,6}+ 
\ket{0,3,5,0}+ 
\ket{3,6,6,1})
\end{align}
The second part of the decoding is to apply \eqref{MMdecode}, which is explicitly evaluated to:

\begin{align}
\label{MMex}
\left( \begin{array}{cc}
M_{k-1}^0(\vec{x}) & M_{k-1}^0(\vec{y_{\overline{\mathcal{J}}}}) 
\end{array}\right)
=
M_2^0(\left\{1,3,6\right\}) = \left(
\begin{array}{ccc}
1&1&1\\
1&3&5\\
1&3^2&5^2
\end{array}\right)
=
\left(
\begin{array}{ccc}
1&1&1\\
1&3&5\\
1&2&4
\end{array}\right).
\end{align}
We verify that applying \eqref{MMex} to the state $\ket{5,1,2,4}$ results in $\ket{1,5,4,4}$ below.
\begin{align}
\nonumber
(5,1,2)
\left(\begin{array}{ccc}
1&1&1\\
1&3&5\\
1&2&4
\end{array}\right)
=
(5+1+2,5+3+4,5+5+8)=(1,5,4)
\end{align}
One can finish verifying that applying \eqref{MMex} to the state $\ket{\psi_{pd}}$ results in:
\begin{align}
\nonumber
\ket{\psi_{pd}}&=\frac{1}{\sqrt{7}} (
\ket{1,5,2,2}+ 
\ket{1,5,3,3}+ 
\ket{1,5,4,4}+ \\
\nonumber
&\indent \indent \indent \ket{1,5,5,5}+ 
\ket{1,5,6,6}+ 
\ket{1,5,0,0}+ 
\ket{1,5,1,1}) \\
&=\frac{1}{\sqrt{7}} \ket{1,5} \otimes \sum_{e\in\mathbf{F}_q}{\ket{e,e}}
\end{align}
The original secret has now been reconstructed in the first $2$ qudits, and the rest of the shares form only a purification of a fully mixed state that is independent of and unentangled with the secret. Therefore the decoding is finished.
\end{example}

\section{Strong Security of the Encoder}
\label{sec4}
In order to prove that the proposed encoding method satisfies the definition of quantum strong security, we prove Condition \ref{l2} of Definition \ref{def1} holds for our encoder. 

%Let $\mathcal{I} \subseteq \{1$, \ldots, $L\}$,
%$\overline{\mathcal{I}} = \{1$, \ldots, $L\} \setminus \mathcal{I}$,
%$\sigma_1 \in \mathcal{S}(\bigotimes_{i \in \mathcal{I}}\mathcal{G}_i)$,
%$\rho_{\mathrm{mix}, \overline{\mathcal{I}}}$ 
%the fully mixed state in $\mathcal{S}(\bigotimes_{i \in \overline{\mathcal{I}}}\mathcal{G}_i)$,
%$\mathcal{J} \subset \{1$, \ldots, $n\}$, and
%$\overline{\mathcal{J}} = \{1$, \ldots, $n\} \setminus \mathcal{J}$.
%If $|\mathcal{J}| \leq k-|I|$,  then
%$\mathrm{Tr}_{\bigotimes_{j \in \overline{\mathcal{J}}}\mathcal{H}_j}W(\sigma_1 \otimes
%\rho_{\mathrm{mix}, \overline{\mathcal{I}}})$ is independent of $\sigma_1$.
%In other words, no quantum information of $\sigma_1$ 
%is leaked to shares whose indices belong to $\mathcal{J}$.

Let $\mathcal{J}\subset\{1,\ldots,n\}$ be the indices of the shares available, and $\mathcal{I}\subset\{1,\ldots,L\}$ be the indices of qudits of the secret, where $|\mathcal{J}|=k-i$ and $|\mathcal{I}|=i$. 
Let the secret on $\mathcal{I}$ be a general pure state
\begin{align}
\ket{\vec{s}_{\mathcal{I}}}=\sum_{\vec{s}\in\mathbf{F}_q^{i}} a_{\vec{s}} \ket{\vec{s}},
\end{align}
where $a_{\vec{s}} \in \mathbb{C}$ and $\sum_{\vec{s}\in\mathbf{F}_q^{L}} |a_{\vec{s}}|^2=1.$
The secret to be encoded is $\rho_{\mathrm{mix},\overline{\mathcal{I}}}\otimes\ket{\vec{s}_{\mathcal{I}}}\bra{\vec{s}_{\mathcal{I}}}$, representing the secret on indices $\mathcal{I}$ and the fully mixed state in $\mathcal{S}(\bigotimes_{i\in\overline{\mathcal{I}}}\mathcal{G}_i)$ on indices $\overline{\mathcal{I}}$. 
We will show that $\mathrm{Tr}_{\bigotimes_{j \in \overline{\mathcal{J}}}\mathcal{H}_j}W(\rho_{\mathrm{mix},\overline{\mathcal{I}}}\otimes\ket{\vec{s}_{\mathcal{I}}}\bra{\vec{s}_{\mathcal{I}}})$ is the fully mixed state in $\mathcal{S}(\bigotimes_{j \in \mathcal{J}}\mathcal{H}_j)$, and therefore independent of $\ket{\vec{s}_{\mathcal{I}}}$.

For ease of exposition, we purify our secret by introducing a reference system of $|\overline{\mathcal{I}}|$ qudits. The purified secret is therefore equal to:
\begin{align}
\ket{\psi_{sec}}=
\left(
\frac{1}{\sqrt{q^{|\overline{\mathcal{I}}|}}}
\sum_{\vec{d} \in \mathbf{F}_q^{L-i}}
\ket{\vec{d}} \otimes \ket{\vec{d}}
\right)
\otimes \ket{\vec{s}_{\mathcal{I}}}.
\label{refsec}
\end{align}
The reference system will henceforth be indexed by $\overline{\mathcal{I}}_2$. 
Let $\mathcal{R}_l$ be $q$-dimensional complex linear spaces, representing qudits on the reference system.

\begin{figure*}
\includegraphics[width=\linewidth]{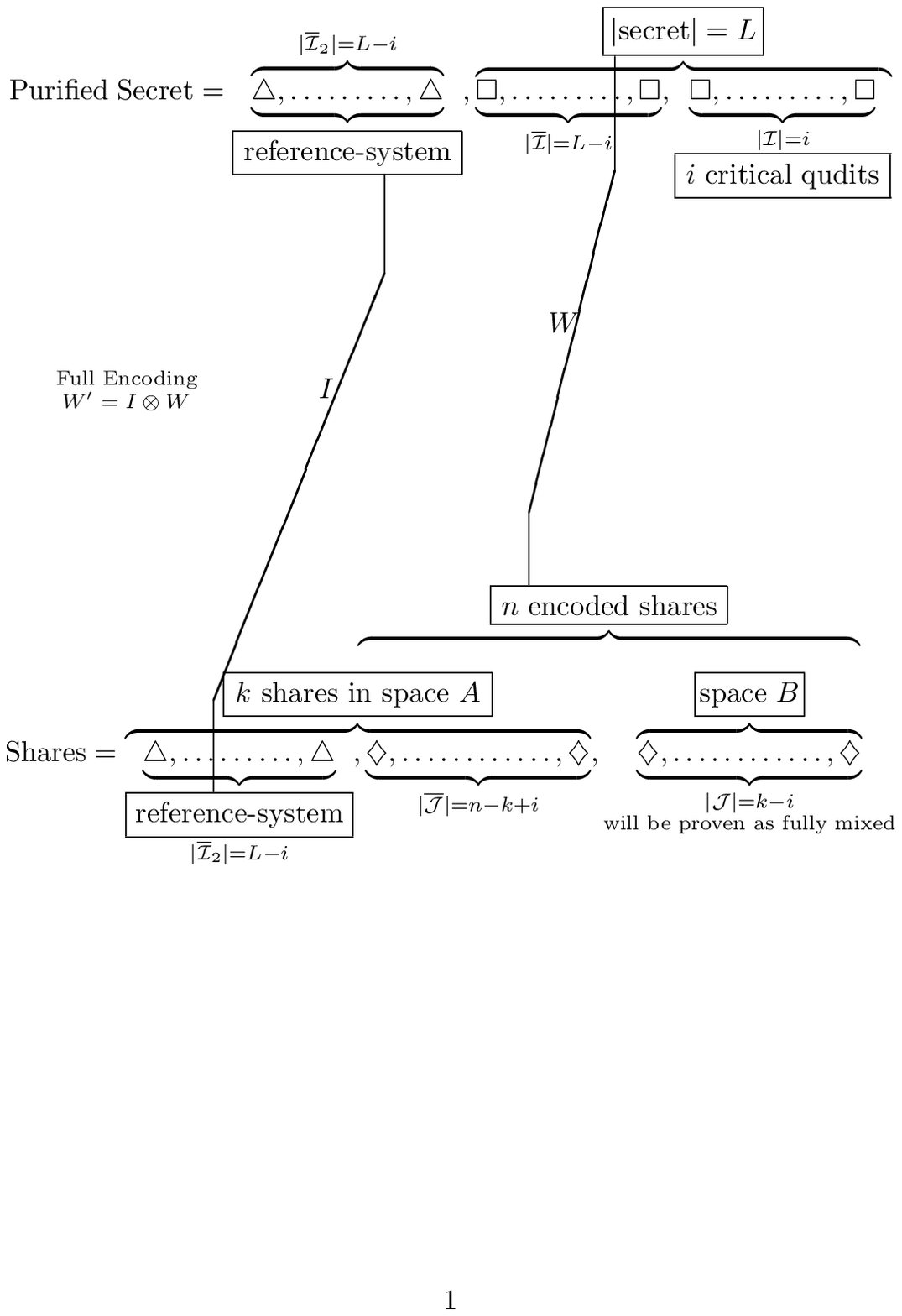}
\caption{%
Strong Security Schematic. $L$ qudits and a reference system of $L-i$ qudits are encoded through $W'$, to result in $n$ shares and the reference system. Recall that $W$ is the encoding map defined in \eqref{encoding}, while $I$ is the identity map. The qudits in space $B = \bigotimes_{j\in\mathcal{J}}\mathcal{H}_j$ are the fully mixed state and have zero information about the $i$ critical qudits of the secret. $\triangle$ represents the reference system qudits, $\Box$ represents the secret qudits, and $\diamondsuit$ represents the encoded shares.
}\label{figure1}
\end{figure*}

We act on the space $\bigotimes_{i\in\left\{1,\ldots,L\right\}}\mathcal{G}_i$ with our encoder. The reference system will remain untouched throughout the proof and be traced out at the end. We denote the $n$ qudits that result from applying our encoder, as ``encoded shares''.
We now define spaces $A =\left(\bigotimes_{l \in \overline{\mathcal{I}}_2}\mathcal{R}_l\right) \otimes \left(\bigotimes_{j\in\overline{\mathcal{J}}}\mathcal{H}_j\right)$
and $B = \bigotimes_{j\in\mathcal{J}}\mathcal{H}_j$. Together, $A$ and $B$ form the space in which the encoded shares and reference system reside.
The entire scheme can be seen in Figure \ref{figure1}. 
%%%%%%%%%%%%%%%%%%%%%%%%%%%%%%%%%%%%%%%%%%%%%%%%%%%%%%%%%%%%%%%%%%%%%%%%%
%%%%%%%%%%%%%%%%%%%%%%%%%%%%%%%%%%%%%%%%%%%%%%%%%%%%%%%%%%%%%%%%%%%%%%%%%
%%%%%%%%%%%%%%%%%%%%%%%%%%%%%%%%%%%%%%%%%%%%%%%%%%%%%%%%%%%%%%%%%%%%%%%%%

Since the reference system is untouched, we can re-express the encoded shares indexed by $\mathcal{J}$ as 
\begin{align}
\nonumber
\mathrm{Tr}&_
{(\bigotimes_{j \in \overline{\mathcal{J}}}\mathcal{H}_j)}
W(\rho_{\mathrm{mix},\overline{\mathcal{I}}}
\otimes
\ket{\vec{s}_{\mathcal{I}}}\bra{\vec{s}_{\mathcal{I}}})
\\\label{TraceChain}
&=
\mathrm{Tr}_
{A}
W'(
\ket{\psi_{sec}} \bra{\psi_{sec}}
),
\end{align}
where $W'=I\otimes W$ is the full encoding. 
%The second step comes from the fact that the reference system can be traced out before or after the encoding without influence to the remaining qudits. This is because the full encoding $W'$ acts on the reference system and the secret qudits separately.

Applying $W'$ to the purified secret $\eqref{refsec}$ results in the transformation from $\ket{\psi_{sec}}$ to:
\begin{align}
\ket{\psi_{enc}}&=
\frac{1}{\sqrt{q^{L-i}}}
\sum_{\vec{s}\in\mathbf{F}_q^i}
a_{\vec{s}}
\left(
\sum_{\vec{d} \in \mathbf{F}_q^{L-i}}
\ket{\vec{d}} \otimes 
\left(
\frac{1}{\sqrt{q^{k-L}}}
\sum_{\vec{c} \in D_k(\vec{d},\vec{s})}
\ket{P_{\vec{c}}(\vec{y})}
\right)\right) 
\nonumber\\
&=\frac{1}{\sqrt{q^{k-i}}}
\sum_{\vec{s}\in\mathbf{F}_q^i}
\sum_{\vec{d} \in \mathbf{F}_q^{L-i}}
\sum_{\vec{c} \in D_k(\vec{d},\vec{s})}
a_{\vec{s}}
\ket{\vec{d}, P_{\vec{c}}(\vec{y})} 
\nonumber\\
&=\frac{1}{\sqrt{q^{k-i}}}
\sum_{\vec{s}\in\mathbf{F}_q^i}
\sum_{\vec{d} \in \mathbf{F}_q^{L-i}}
\sum_{\vec{c} \in D_k(\vec{d},\vec{s})}
a_{\vec{s}}
\ket{P_{\vec{c}}(\vec{x_{\overline{\mathcal{I}}}}), P_{\vec{c}}(\vec{y})}, 
\label{encsec}
\end{align}
where $(\vec{d},\vec{s})$ signifies the horizontal concatenation of $\vec{d}$, and $\vec{s}$, creating one vector of length $L$. 
The second step in \eqref{encsec} comes from the definition of 
$D_k(\vec{v}=(v_1,\ldots,v_L))$ as the set of polynomial coefficients such that the 
evaluation of the polynomial on $\vec{x}$ is equal to $\vec{v}$.

We now re-express \eqref{encsec} into the following:
\begin{align}
\nonumber
\ket{\psi_{enc}}&=
\frac{1}{\sqrt{q^{k-i}}}
\sum_{\vec{s}\in\mathbf{F}_q^i}
\sum_{\vec{d} \in \mathbf{F}_q^{L-i}}
\sum_{\vec{c} \in D_k(\vec{d},\vec{s})}
a_{\vec{s}}
\ket{P_{\vec{c}}(\vec{x_{\overline{\mathcal{I}}}}), P_{\vec{c}}(\vec{y}_{\overline{\mathcal{J}}}), P_{\vec{c}}(\vec{y}_{\mathcal{J}})}
\\\label{van-1}
&=
\frac{1}{\sqrt{q^{k-i}}}
\sum_{\vec{g}\in\mathbf{F}_q^L}
\sum_{\vec{c} \in D_k(\vec{g})}
a_{\vec{g}_{\mathcal{I}}}
\ket{P_{\vec{c}}(\vec{x_{\overline{\mathcal{I}}}}), P_{\vec{c}}(\vec{y}_{\overline{\mathcal{J}}}), P_{\vec{c}}(\vec{y}_{\mathcal{J}})},
%\frac{1}{\sqrt{q^{k-i}}}
%\sum_{\vec{c} \in E_k(\vec{s}_{\mathcal{I}})}
%\ket{P_{\vec{c}}(\vec{x_{\overline{\mathcal{I}}}}), P_{\vec{c}}(\vec{y}_{\overline{\mathcal{J}}}), P_{\vec{c}}(\vec{y}_{\mathcal{J}})},
\end{align}
%where $E_k(\vec{s}_{\mathcal{I}})=\{\vec{c} \mid P_{\vec{c}}(\vec{x_{\mathcal{I}}})=\vec{s}_{\mathcal{I}}\}$.
where $\vec{g}=(\vec{d},\vec{s})$, $\vec{g}_{\mathcal{I}}=\vec{s}$, and 
$\vec{g}_{\overline{\mathcal{I}}}=\vec{d}$.
Note that the number of shares in space $A$ is equal to 
the length of the vector
$(P_{\vec{c}}(\vec{x_{\overline{\mathcal{I}}}}), 
P_{\vec{c}}(\vec{y}_{\overline{\mathcal{J}}}))$.
This length is $|\overline{\mathcal{I}}|+|\overline{\mathcal{J}}|=(L-i)+(n-k+i)$. Since $n-k=k-L$, the number of shares in space $A$ are $L-i+k-L+i=k.$ 
Therefore $(P_{\vec{c}}(\vec{x_{\overline{\mathcal{I}}}}), 
P_{\vec{c}}(\vec{y}_{\overline{\mathcal{J}}}))$
is a vector of $k$ evaluations of the polynomial, 
specified by $\vec{c}$, on pairwise distinct elements of 
$\mathbf{F}_q$. By Lemma \ref{lem1} this uniquely 
specifies the polynomial, and therefore $\vec{c}$ as well. 

We can now calculate the density operator 
$\rho_{enc}=\ket{\psi_{enc}}\bra{\psi_{enc}}$ in \eqref{density-operator}. 
\begin{align}
\label{density-operator}
\rho_{enc} = 
\frac{1}{q^{k-i}}
\sum_{\vec{h},\vec{g}\in\mathbf{F}_q^L}
%\sum_{\substack{\text{$\vec{c}_1 \in D_k(\vec{g})$}
%\\ 
%text{$\vec{c}_2 \in D_k(\vec{h})$}}}
\sum_{\substack{\vec{c}_1 \in D_k(\vec{g})
\\ 
\vec{c}_2 \in D_k(\vec{h})
}}
\begin{array}{r}
\\
a_{\vec{g}_{\mathcal{I}}}
a_{\vec{h}_{\mathcal{I}}}^\dag
%%%%%%%%%
\ket{ P_{\vec{c}_1}(\vec{x_{\overline{\mathcal{I}}}},
%), P_{\vec{c}_1}(
\vec{y}_{\overline{\mathcal{J}}})}
\bra{P_{\vec{c}_2}(\vec{x_{\overline{\mathcal{I}}}},
%), P_{\vec{c}_2}(
\vec{y}_{\overline{\mathcal{J}}})}
\\
\otimes
\ket{P_{\vec{c}_1}(\vec{y}_{\mathcal{J}})}
\bra{P_{\vec{c}_2}(\vec{y}_{\mathcal{J}})}
\end{array}
\end{align}
%%%%%%%%%%%%%%%%%%%%%%%%%%%%%%%%%%%%%%%%%%%%%%%%%%%%%%%%%%%%%%%%%%%%%%
The notation $\ket{ P_{\vec{c}_1}(\vec{x_{\overline{\mathcal{I}}}}), P_{\vec{c}_1}(
\vec{y}_{\overline{\mathcal{J}}})}$ is written as
$\ket{ P_{\vec{c}_1}(\vec{x_{\overline{\mathcal{I}}}},\vec{y}_{\overline{\mathcal{J}}})}$
for visibility purposes.
Now we perform the partial trace over space $A$ on $\rho_{enc}$ to obtain
\begin{align}
\nonumber
\Tr_A [\rho_{enc}] 
&=
\frac{1}{q^{k-i}}
\sum_{\vec{h},\vec{g}\in\mathbf{F}_q^L}
a_{\vec{g}_{\mathcal{I}}}
a_{\vec{h}_{\mathcal{I}}}^\dag
\sum_{\substack{\vec{c}_1 \in D_k(\vec{g})
\\ 
\vec{c}_2 \in D_k(\vec{h})
}}
\begin{array}{r}
\\
\overbrace{\Tr \left[
\ket{ P_{\vec{c}_1}(\vec{x_{\overline{\mathcal{I}}}},
\vec{y}_{\overline{\mathcal{J}}})}
\bra{P_{\vec{c}_2}(\vec{x_{\overline{\mathcal{I}}}},
\vec{y}_{\overline{\mathcal{J}}})}
\right] }^{=\delta_{\vec{c}_1,\vec{c}_2}}
\\
\ket{P_{\vec{c}_1}(\vec{y}_{\mathcal{J}})}
\bra{P_{\vec{c}_2}(\vec{y}_{\mathcal{J}})}
\end{array}
\\
&=
\frac{1}{q^{k-i}}
\sum_{\vec{h},\vec{g}\in\mathbf{F}_q^L}
a_{\vec{g}_{\mathcal{I}}}
a_{\vec{h}_{\mathcal{I}}}^\dag
\sum_{\substack{\vec{c}_1 \in D_k(\vec{g})
\\ 
\vec{c}_2 \in D_k(\vec{h})
}}
\delta_{\vec{c}_1,\vec{c}_2}
\ket{P_{\vec{c}_1}(\vec{y}_{\mathcal{J}})}
\bra{P_{\vec{c}_2}(\vec{y}_{\mathcal{J}})}.
\label{traced1}
\end{align}
The trace evaluates to the delta function because as noted earlier, the space $A$ contains $k$ evaluations 
of a polynomial, which, by Lemma \ref{lem1}, specifies $\vec{c}$ its coefficients.

Note that for all $\vec{g},\vec{h}\in\mathbf{F}_q^L$, the intersection of $D_k(\vec{g})$ 
and $D_k(\vec{h})$ 
is empty if and only if $\vec{g}\neq\vec{h}$. 
The traced state \eqref{traced1} is therefore equal to
\begin{align}
\nonumber
\Tr_A[\rho_{enc}]&=
\frac{1}{q^{k-i}}
\sum_{\vec{g}\in\mathbf{F}_q^L}
|a_{\vec{g}_{\mathcal{I}}}|^2
\sum_{\vec{c} \in D_k(\vec{g})}
\ket{P_{\vec{c}}(\vec{y}_{\mathcal{J}})}
\bra{P_{\vec{c}}(\vec{y}_{\mathcal{J}})}
\\\nonumber&=
\frac{1}{q^{k-i}}
\sum_{\vec{s}\in\mathbf{F}_q^i}
|a_{\vec{s}}|^2
\sum_{\vec{d}\in\mathbf{F}_q^{L-i}}
\sum_{\vec{c} \in D_k(\vec{d},\vec{s})}
\ket{P_{\vec{c}}(\vec{y}_{\mathcal{J}})}
\bra{P_{\vec{c}}(\vec{y}_{\mathcal{J}})}
\\
&=
\frac{1}{q^{k-i}}
\sum_{\vec{s}\in\mathbf{F}_q^i}
|a_{\vec{s}}|^2
\sum_{\vec{c} \in E_k(\vec{s})}
\ket{P_{\vec{c}}(\vec{y}_{\mathcal{J}})}
\bra{P_{\vec{c}}(\vec{y}_{\mathcal{J}})},
\label{traced2}
\end{align}
where $E_k(\vec{s}) = 
\left\{
\vec{c} \in \mathbf{F}_q^k 
\mid 
P_{\vec{c}}(\vec{x}_{\mathcal{I}})
=
\vec{s} 
\right\}$.
By the same argument that equates \eqref{edecode} to \eqref{edecode2}, \eqref{traced2} is
equivalent to 
\begin{align}
\Tr_A[\rho_{enc}]&=
\frac{1}{q^{k-i}}
\sum_{\vec{s}\in\mathbf{F}_q^i}
|a_{\vec{s}}|^2
\sum_{\vec{v} \in \mathbf{F}_q^{k-i}}
\ket{\vec{v}}\bra{\vec{v}}
\nonumber
\\
\label{van-2}
&=
\frac{1}{q^{k-i}}
\sum_{\vec{v} \in \mathbf{F}_q^{k-i}}
\ket{\vec{v}}\bra{\vec{v}},
\end{align}
which can easily be seen as the fully mixed state in space B. Since this subsystem is the fully mixed state, it must be independent of the secret $\ket{\vec{s}_{\mathcal{I}}}$. Thus Condition \ref{l2} is satisfied, and our encoder has quantum strong security.

\begin{example}
We now provide a concrete example of the strong security in a ($k=3,L=2,n=4$) quantum ramp secret sharing scheme. We retain the notations in this section. Let $q=7$ be the dimension of each qudit. The public values $\vec{x}$ and $\vec{y}$ are $\vec{x}=\left(1,3\right)$, and $\vec{y}=\left(6,2,4,5\right)$. Let $\mathcal{I}=\left\{2\right\}$, be the index of the qudit of the secret that an unqualified set of participants wants to steal. Let the set of unqualified participants be $\mathcal{J}=\left\{3,4\right\}$. By the strong security condition, the shares $3$ and $4$ must be unable to produce any information on the second qudit. Let the second qudit be $\ket{s_2=5}$. The full secret, including the reference system is therefore
\begin{align*}
%\label{pure-sec}
\ket{\psi_{\mathrm{pure}}}=
\frac{1}{\sqrt{7}}
\left(
\ket{0, 0, 5} +
\ket{1, 1, 5} +
\ket{2, 2, 5} +
\ket{3, 3, 5} +
\ket{4, 4, 5} +
\ket{5, 5, 5} +
\ket{6, 6, 5}
\right).
\end{align*}

By applying the proposed encoding, we arrive at a state which is a superposition of $q^2=49$ basis states. We explicity show $7$ basis states out of $49$, which come from encoding $\ket{1,1,5}$. Recall that the encoder only acts on $L=2$ qudits, which in this case is the last two qudits $\ket{1,5}$. We have already shown the encoded state of this is $\ket{\psi_{ex}}$ in \eqref{example-encode}. Therefore, the encoded state is
\begin{align}
\nonumber
\ket{1}\otimes\ket{\psi_{ex}},
\end{align}
a superposition of $7$ basis states. One can verify that the entire superposition of $49$ basis states is
\footnotesize
\begin{align*}
%\label{LARGE}
&\hphantom{\;\;\;\;}\ket{\psi_{sec-enc}} = \\
&\frac{\ket{0}}{7} \otimes \left( \ket{2,6,4,3}+\ket{3,5,0,4}+\ket{4,4,3,5}+\ket{5,3,6,6}+\ket{6,2,2,0}+\ket{0,1,5,1}+\ket{1,0,1,2} \right) +\\
&\frac{\ket{1}}{7} \otimes \left(\ket{\psi_{ex}}\right) +\\
&\frac{\ket{2}}{7} \otimes \left( \ket{6,0,3,1}+\ket{0,6,6,2}+\ket{1,5,2,3}+\ket{2,4,5,4}+\ket{3,3,1,5}+\ket{4,2,4,6}+\ket{5,1,0,0} \right) +\\
&\frac{\ket{3}}{7} \otimes \left( \ket{1,4,6,0}+\ket{2,3,2,1}+\ket{3,2,5,2}+\ket{4,1,1,3}+\ket{5,0,4,4}+\ket{6,6,0,5}+\ket{0,5,3,6} \right) +\\
&\frac{\ket{4}}{7} \otimes \left( \ket{3,1,2,6}+\ket{4,0,5,0}+\ket{5,6,1,1}+\ket{6,5,4,2}+\ket{0,4,0,3}+\ket{1,3,3,4}+\ket{2,2,6,5} \right) +\\
&\frac{\ket{5}}{7} \otimes \left( \ket{5,5,5,5}+\ket{6,4,1,6}+\ket{0,3,4,0}+\ket{1,2,0,1}+\ket{2,1,3,2}+\ket{3,0,6,3}+\ket{4,6,2,4} \right) +\\
&\frac{\ket{6}}{7} \otimes \left( \ket{0,2,1,4}+\ket{1,1,4,5}+\ket{2,0,0,6}+\ket{3,6,3,0}+\ket{4,5,6,1}+\ket{5,4,2,2}+\ket{6,3,5,3} \right).
\end{align*}
\normalsize

As in Section \ref{sec4}, we now show that the subsystem in space Y of the shares $\ket{\psi_{sec-enc}}\bra{\psi_{sec-enc}}$ is the fully mixed state. This is equivalent to showing that the last two qudits are fully mixed. It can be observed that the basis indices of any $3$ qudits are unique. For our case, let the first three qudits be traced out. We can verify that the remaining $2$ qudits have basis indices filling the entire set $\left\{(0,0),(0,1),\ldots,(6,6)\right\}$. Thus the qudits in space Y are fully mixed. In other words, they have no information about the secret $\ket{\vec{s}_{\mathcal{I}}}$.
\end{example}

\section{Conclusion}
\label{sec6}
In this paper, we have shown that Ogawa et al.'s secret sharing scheme does not satisfy the strong security condition for quantum ramp secret sharing. We have provided a quantum strongly secure ramp secret sharing scheme based on its classical analog. The difference between this encoding and Ogawa et al.'s encoding, is that the secret is encoded into evaluations of polynomials instead of their coefficients. We have provided a decoding method so that, given any $k$ shares, the secret can be reconstructed. Finally we proved that this encoding method is strongly secure 
by showing that any $k-i$ encoded shares form a fully mixed state. Numerical examples are provided for each proof. The coding efficiency of our encoding is the same as that of the conventional quantum ramp SSs in \cite{ogawa05}.

\begin{acknowledgements}
This research was conducted as part of
Tokyo Institute of Technology International Research Opportunities Program
under
Re-Inventing Japan Project
funded by
Ministry of Education, Culture, Sports, Science and Technology.

This research is partly supported by the National
Institute of Information and Communications Technology,
Japan, and by the Japan Society for the Promotion of Science Grant
 Nos.\ 23246071 and 26289116.
\end{acknowledgements}

% Non-BibTeX users please use

\end{document}